\documentstyle[preprint,epsf,aps]{revtex}
\begin{document}
\tightenlines
\draft
\title{Microscopic study of freeze-out in relativistic heavy ion 
       collisions at SPS energies}
\author{  L.V. Bravina,$^{1,2}$ 
          I.N. Mishustin,$^{3,4}$
       \footnote{Present address: Institut f\"ur Theoretische 
       Physik, Goethe Universit\"at Frankfurt, Germany}
          J.P. Bondorf,$^3$ Amand Faessler,$^1$
          E.E. Zabrodin$^{1,2}$
 }
\address{
$^1$ Institut f\"ur Theoretische Physik, Universit\"at T{\"u}bingen, 
Auf der Morgenstelle 14, D-72076 T{\"u}bingen, Germany \\
$^2$ Institute for Nuclear Physics, Moscow State University,
 119899 Moscow, Russia \\
$^3$ The Niels Bohr Institute, Blegdamsvej 17, DK-2100, Copenhagen O, 
 Denmark \\
$^4$ The Kurchatov Institute, Russian Scientific Center, 123182 
 Moscow, Russia \\
}

\maketitle

\begin{abstract}
The freeze-out conditions in the light (S+S) and heavy (Pb+Pb)
colliding systems of heavy nuclei at 160 AGeV/$c$ are analyzed within 
the microscopic Quark Gluon String Model (QGSM).
We found that even for the most heavy systems particle emission takes 
place from the whole space-time domain available for the system
evolution, but not from the thin ''freeze-out hypersurface", adopted 
in fluid dynamical models. Pions are continuously emitted from the 
whole volume of the reaction and reflect the main trends of the 
system evolution. Nucleons in Pb+Pb collisions initially come from 
the surface region. For both systems there is a separation of the
elastic and inelastic freeze-out. The mesons with large  transverse 
momenta, $p_t$, are predominantly produced at the early stages of the 
reaction. The low $p_t$-component is populated by mesons coming mainly 
from the decay of resonances. This explains naturally the decreasing 
source sizes with increasing $p_t$, observed in HBT interferometry. 
Comparison with S+S and Au+Au systems at 11.6 AGeV/$c$ is also 
presented.
\end{abstract}

\widetext

\section{Introduction}
\label{sec1}

The main goal of the high energy heavy-ion experiments at Brookhaven 
and CERN is to study properties of hot and dense hadronic matter 
produced in the course of the nuclear collisions. Using the measured 
final state distributions one tries to reconstruct the dynamical 
picture of the nuclear reaction and recognize the new phenomena 
associated with possible quark-hadron phase transition, like 
Quark-Gluon Plasma (QGP) formation. This can be done by comparison of
the experimental data with the predictions of different models. The 
reaction dynamics and experimental constraints are so complicated at 
high energies that any single model cannot succeed to reproduce the 
whole set of data. But the discrepancies between the standard models 
and the data may help to reveal new phenomena. The recent Pb+Pb 
experiments at 160 AGeV/$c$ at CERN try to understand whether there 
is new physics at these high energies in such heavy colliding systems.

There are several groups of models aiming to describe heavy ion 
collisions at relativistic energies - microscopic string models 
\cite{ABS90,ABCTGS93,URQMD,RQMD,ARC,ART}, macroscopic fluid 
dynamical- \cite{BMGR,Strott} and thermal models 
\cite{Heinz,Stachel95,Goren97}.
The macroscopic models use different {\it ad hoc\/} assumptions 
concerning the freeze-out stage when particle spectra are calculated
for completing for example the fluid dynamical equations, while in 
microscopic models the freeze-out conditions are obtained 
automatically without special efforts. A direct comparison of 
microscopic models with hydrodynamical models is done, e.g., in 
\cite{BCLS94,BACLS94,Shuryak}.

In present paper we continue to study the freeze-out conditions in 
relativistic heavy ion collisions which were initiated earlier 
\cite{BCLS94,Freeze,Freeze2}.
The calculations are carried out within the Quark-Gluon String Model
(QGSM) \cite{ABS90,ABCTGS93} for light (S+S) and heavy (Pb+Pb) systems 
at 160 AGeV. Below we present the distributions of coordinates and 
momenta of final state hadrons averaged over the whole ensemble of 
events. Therefore, we study the one-body characteristics of the 
reactions disregarding the fluctuations on the level of single events.
One should keep this in mind when comparing results with emitting 
source parameters extracted from the two-body correlation functions in 
Hanbury-Brown-Twiss (HBT) interferometric measurements \cite{Wein99}. 
The similar analysis has been done also within the RQMD 
\cite{Sorge9697,Matti9597} and the UrQMD \cite{Bass} models.

\section{Quark-Gluon String Model} 
\label{sec2}

\subsection{The main features of the model}
\label{s2sub1}

The QGSM is based on the Regge- and string phenomenology of particle 
production in inelastic hadron-hadron collisions at high energies 
\cite{QGSM,QGSMhh}. In application to hadron-nucleus and 
nucleus-nucleus  collisions this picture was supplemented by a proper 
treatment of the multiple secondary interactions of hadrons (cascades) 
\cite{QGSMaa}. The model incorporates the string fragmentation, 
resonance formation and hadron rescatterings. To keep the model simple 
we did not consider here any modifications like color fluctuations, 
mean fields, enhanced cross sections or string fusion. 
Even without these modifications QGSM reproduces the main features of
hadronic and nuclear collisions rather good. 

In addition to pions, $\pi$, and nucleons, $N$, QGSM takes into 
account vector mesons $(\rho,\omega)$ and low-lying baryon 
resonances,  mostly $\Delta(1232)$. Strange particles (K, $\Lambda$, 
$\Sigma$) are also included in the model. Within the QGSM the pion 
absorption is described by a two-step mechanism, including 
$\pi N\!\rightarrow \!\Delta$ and $\Delta N\!\rightarrow \!NN$ 
reactions. The model describes particle spectra in hadron-hadron 
({\it hh\/}) \cite{ABS90,AM91}, hadron-nucleus ({\it hA\/}) and 
nucleus-nucleus ({\it AA\/}) collisions \cite{NA35} quite well 
but underestimates by a factor of 2-4 the production of strange 
particles in the case of S+S collisions at SPS energies 
\cite{ABCTGS93}. The model can be improved by decreasing the formation 
time of the produced hadrons, by including higher mass baryon 
resonances and the 3-body channel of pion absorption, $\pi NN\!
\rightarrow \!NN$, and by increasing the pion absorption cross 
section, $\sigma {_{\Delta N\rightarrow NN}}$, in dense baryon medium. 

The positions and momenta of nucleons inside the nuclei are generated 
according to the Woods-Saxon density distribution and the Fermi 
momentum distribution, respectively. Two nucleons with four-momenta  
$p_1$ and $p_2$ and total center-of-mass (CM) energy squared, 
$s=(p_1+p_2)^2$, can interact inelastically with the cross section 
$\sigma _{inel}(s)$, or elastically with the cross section 
$\sigma _{elas}(s)$, when they approach each other closer than 
$\sqrt{\sigma / \pi }$. Here $\sigma=\sigma _{inel}+\sigma _{elas}$ 
is the total cross section  (the so-called ``black disk" 
approximation). Pauli blocking is taken into account by excluding 
of the already occupied final states from the available phase space.

The primary and subsequent interactions of hadrons are simulated by 
the Monte-Carlo method according to their momenta and positions. 
In primary inelastic collisions the particles appear mainly through 
the formation and fragmentation of strings. Due to the uncertainty 
principle hadrons are formed and can interact further only after some 
formation time. Those hadrons which contain the valence quarks of the 
colliding hadrons can interact immediately with the cross section 
$\sigma _{qN}$, taken from the additive quark model \cite{aqm}.
Angular and momentum distributions of secondary particles produced 
in elementary {\it hh\/} collisions are adjusted to available 
experimental data \cite{Blobel,NA22}. In case of lack of the 
experimental cross sections the one-pion exchange model, detailed 
balance considerations and isospin symmetry arguments are used.
Since the particles loose energy in the course of their 
rescatterings, after some time they can either interact only 
elastically or be produced via the resonance decays.  Finally, all 
interactions and decays cease. This stage corresponds to the 
experimental freeze-out conditions.

In the present paper we compare S+S and Pb+Pb collisions at SPS 
bombarding momentum, $p_{lab} = 160$ GeV/$c$ per nucleon. This 
corresponds to $\sqrt{s}=17.4$ GeV, $v_{cm} = 0.9942$, 
$\gamma_{cm} = 9.26$ and CM rapidity $y_{cm} = 2.916$. The central 
{\it AA\/} collisions are traced in the CM frame beginning from the
sufficiently early time $t=0$ which has been chosen at the moment when 
the nuclear centers are separated by two nuclear radii, $2R_A$. The 
projectile and target nuclei, initially Lorentz contracted by 
$\gamma_{cm}$, are propagating with velocities $\pm v_{cm}$. Due to
the Lorentz contraction of nuclei, the first nucleon-nucleon collision 
occurs at time $t_0\approx 4.85$ fm/$c$ in S+S collisions 
($R_{S}=3.56$ fm) and $t_0\approx 7.40$ fm/$c$ in Pb+Pb collisions 
($R_{Pb}=6.64$ fm). Our analysis is based on the statistics of 1000 
S+S and 50 Pb+Pb very central events ($b=0.1$ fm). The calculations 
are stopped at $t_f=50$ fm/$c$ when most of the particles are already
in their final states.

\subsection{Global kinetic characteristics}
\label{s2sub2}

It is interesting to see the model predictions for different channels 
of elementary interactions listed in Table~\ref{tab1}. According to 
QGSM the number of elementary interactions in Pb+Pb (S+S) collisions 
is about 18 500 (410), which is two times larger than that in Au+Au 
(S+S) collisions at $p_{lab} = 11.6$ AGeV/$c$. Among them there are 
6500 (200) inelastic and 11500 (200) elastic collisions plus 2000 
(116) decays of resonances. In Pb+Pb (S+S) central collisions 416 
(58) baryons (including the initial ones) and 3830 (340) mesons are 
produced. Combining these numbers one can estimate the mean number of
collisions per hadron for Pb+Pb collisions:  18500/(4200/2)=9 and for 
S+S collisions: 410/(400/2)=2. Similar results were obtained within 
RQMD for Pb+Pb and S+S collisions at 200 AGeV \cite{Keitz91}.
Certainly, just a couple of collisions per hadron, such as in the case 
of S+S interactions, is not large enough to reach the stage of local 
thermal and chemical equilibration. The situation for Pb+Pb looks more 
promising. The detailed analysis of particle energy spectra and 
abundances has been done in UrQMD \cite{lv98plb,lv99jpg,lep1} for the 
central cell of heavy ion collisions at energies spanning from AGS
up to SPS. It was found that despite reaching local kinetic
equilibrium the hadronic matter even in the central cell is still
far from the stage of thermal and chemical equilibrium at SPS 
energies. However, the study of the relaxation of hot hadronic
matter to full equilibrium in the QGSM model lies out of scope of the
present paper and will be done in forthcoming publications
\cite{eqQGSM}.

\section{Space-time freeze-out picture}
\label{sec3}

Here we study separately the last interaction points of the nucleons 
and pions produced in inelastic and in elastic collisions, as well as 
in resonance decays. It is worth noting that inelastic collisions are 
responsible for chemical equilibration of the system, while elastic 
collisions drive the system towards the thermal equilibration. 
The resonance decays characterize mostly the individual properties of 
the emitted particles. The long-living resonances carry little 
information about the reaction zone.  

The phase-space distribution for the particles on the mass shell is a 
function of seven independent variables: $({\vec r}$, ${\vec p}$, 
$t)$. It is quite difficult to imagine this manyfold in 
eight-dimensional space. For the sake of simplicity we integrate it 
over some variables and study separately different space-time and 
phase-space three dimensional distributions. In the case of the 
collision of symmetric nuclei, where two coordinates in the transverse 
plane $(x,y)$ as well as $(p_x,p_y)$ are equivalent, there are only 10 
different coordinate pairs: $(t,z)$, $(t,r_T)$, $(t,p_z)$, $(t,p_T)$, 
$(z,r_T)$, $(z,p_z)$, $(z,p_T)$, $(r_T,p_z)$, $(r_T,p_T)$, and 
$(p_z,p_T)$. Below we show almost all distributions in these planes,
namely $d^2N/dzdt$, $d^2N/r_Tdr_Tdt$, $d^2N/dy_{cm}dt$, 
$d^2N/m_Tdm_Tdt$ and their projections on the
$t$-, $z$-, $r_T$-, $y$- and $m_T$-axes.

\subsection{Longitudinal and transverse directions}
\label{s3sub1}

\begin{figure}[htp]
\centerline{\epsfysize=13cm \epsfbox{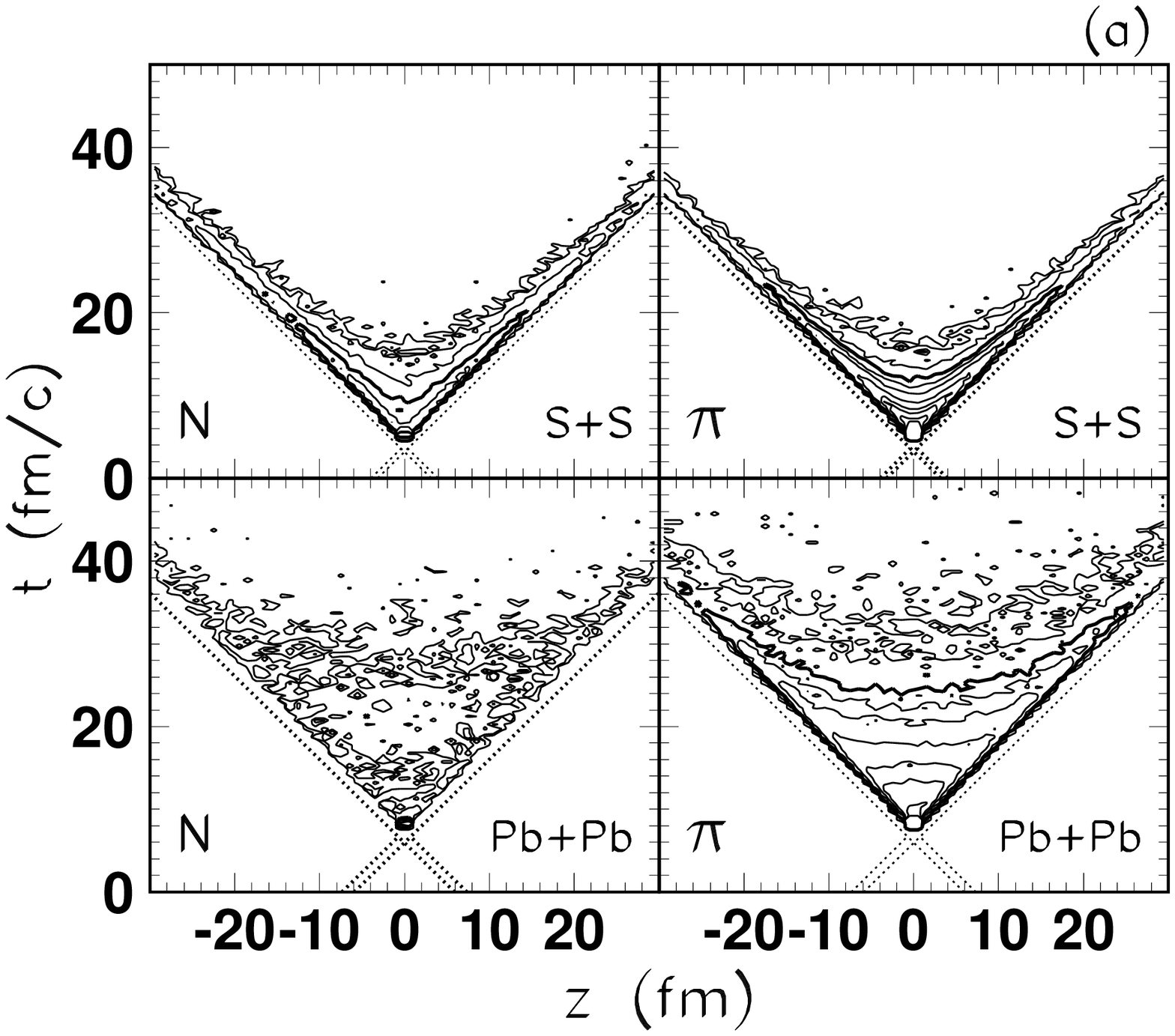}}
\caption{
{\bf (a)}: $d^2N/dzdt$ distribution of the final state hadrons
over $(t,z)$- coordinates of their last elastic and inelastic 
collision points. Distributions are presented separately for nucleons
(left panels) and pions (right panels), produced in S+S (upper row) and 
Pb+Pb (lower row) central (b=0.2 fm) collisions at 160  AGeV/$c$. 
Contour plots correspond to 
$d^2N/dzdt=$0.005, 0.01, 0.033, 0.1, 0.2, 0.4, 1.0, 2.0, 3.0, 5.0, 7.0 
particles/fm$^2$/$c$ for S+S collisions, 
$d^2N/dzdt=$0.1, 0.2, 0.7, 1.0, 2.0, 3.0, 5.0, 7.0, 9.0, 10.0 
nucleons/fm$^2$/$c$ 
and $d^2N/dzdt=$0.06, 0.2, 0.7, 1.3, 2.6, 6.5, 13.0, 20.0, 33.0, 43.0 
pions/fm$^2$/$c$ for Pb+Pb collisions. The contours with 
$d^2N/dzdt=0.1$ (0.65) are thickened for S+S (Pb+Pb) reactions, 
respectively. The dotted lines show the trajectories of the nuclear 
edges.\\
{\bf (b)}: $d^2N/\eta d\eta \tau d\tau$- distribution of the final 
state pions over $(\tau ,\eta )$- coordinates of their last collision 
points. 
The sequence of the contour plots is the same as in {\bf (a)}.
}
\centerline{\epsfysize=15cm \epsfbox{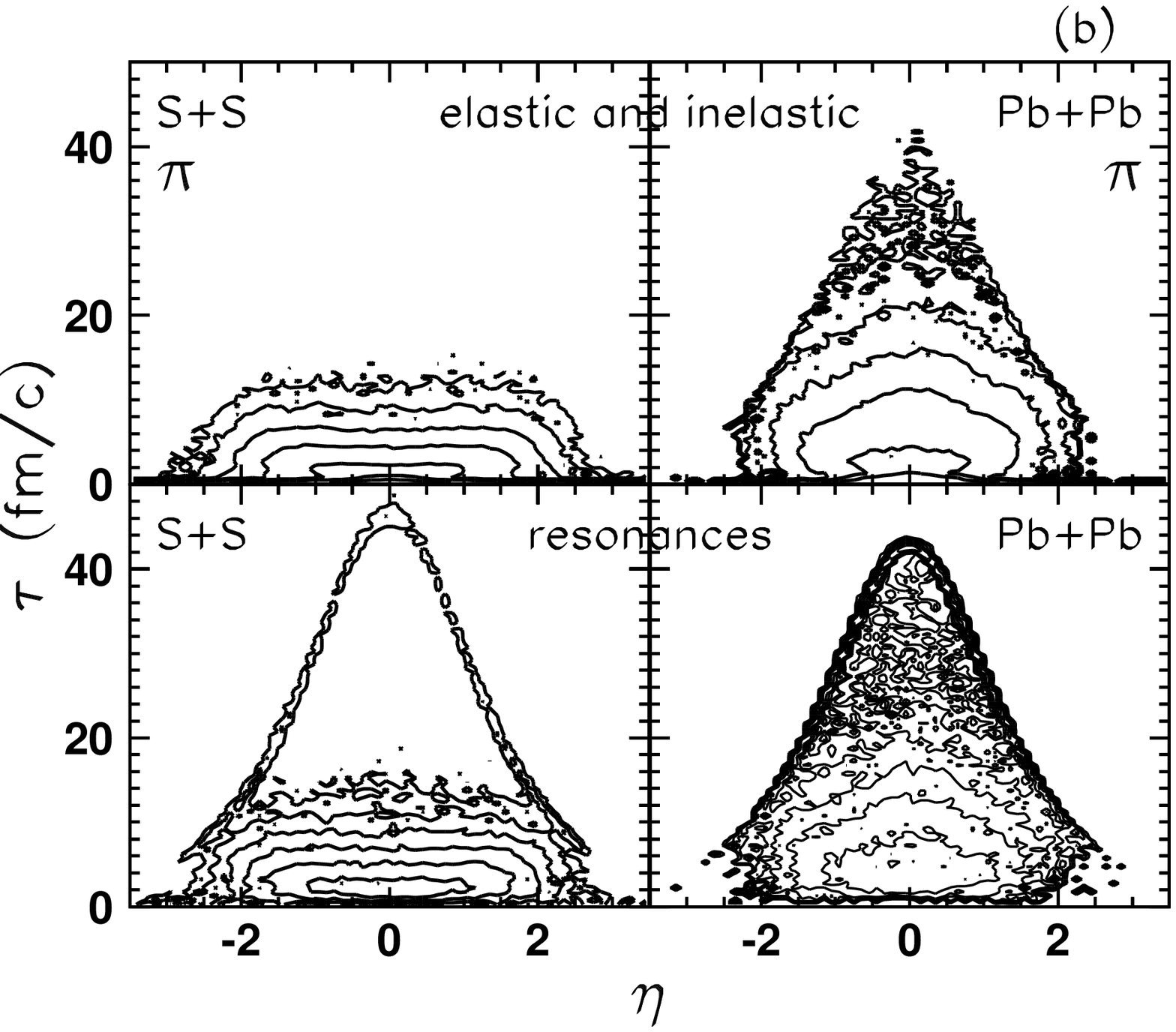}}
\label{fig1}
\end{figure}

Figure~\ref{fig1}(a) depicts the distribution of the emitted nucleons 
and pions over longitudinal coordinate and time. Both for light (S+S) 
and for heavy (Pb+Pb) colliding systems the distribution of the final 
state hadrons, ${d^2N}/{dtdz}$, over the $(t,z)$-coordinates of 
their last interactions shows that the particles are emitted from
the whole available space-time region. In this respect the freeze-out 
picture obtained in QGSM is different from that of the Landau's or 
Bjorken's models \cite{Land53,CF74,Bjor83}, based on sharp freeze-out. 
Nevertheless, the shape of the contours is concave and  similar to the 
Bjorken proper-time surface. This is partially due to the fact that 
the transverse spacial coordinates ($x,y$) are integrated over.

For all cases the QGSM predicts a high narrow peak coming from the 
beginning of the reaction (early emitted particles). 
However, the integrated number of particles coming from the peak is 
not so high: there are about 200 pions and 40 nucleons for the heavy
Pb+Pb system. 

Later on the emissivity drops gradually with increasing $t$ and $z$.
The emission region spreads throughout the whole region inside the 
lightcone. The picture is very different for nucleons and pions in 
Pb+Pb collisions. In addition to the narrow high peak at small values 
$(z,t)$ for nucleons, there appears a broad and flat maximum at 
$z=\pm 5$ and $t=17\div 28$ fm/$c$. This plateau corresponds to the 
''thermal" component of the nucleon distribution due to many elastic 
and inelastic collisions.

In contrast to S+S collisions, much stronger nucleon stopping is
observed in Pb+Pb reactions. For nucleons in Pb+Pb collisions the 
distribution deviates significantly from the lightcone, which would 
result from the full transparency. Still, the longitudinal expansion 
for both nucleons and pions is considerable. The emission zone is 
much wider than the initial longitudinal size of Lorentz contracted 
nuclei. Using the distributions in $(\eta,\tau)$-plane one can 
estimate the longitudinal expansion velocity. In Pb+Pb the 
distribution is narrower compared to that of S+S reflecting again 
the larger transparency and stronger longitudinal motion in latter 
case. In Pb+Pb the concept of collective flow is more justified due to 
the higher degree of equilibration. Here the matter expands with the 
collective longitudinal velocity $v_{cm}=0.944$, corresponding to the 
space-time pseudorapidity $\eta=1/2\ln{\left[ (t+z)/(t-z) \right]} =
1.7$. It is much smaller than the initial CM rapidity of projectile 
and target, $y_{cm}=2.9$. The particles from the initial high peak in 
$d^2N/dzdt$-distribution are emitted mostly from the outer edge 
surface of the colliding nuclei. 

This picture is supported by the $d^2N/d\eta \tau d\tau$-distribution  
of pions over the variables ($\eta$,$\tau$) shown in 
Fig.~\ref{fig1}(b), where $\eta$ is the pseudorapidity and 
$\tau=\sqrt{(t-t_0)^2-z^2}$  is a proper time. 
In these variables the sharp Bjorken freeze-out would look like 
hypersurface $\tau = const$. 
Only in S+S collisions it has some similarity with that picture,
but anyway the region of emission is wide in $\tau $-direction. 
In contrast to simple picture the emission proceeds during the whole
period  $0 \leq \tau \leq 15$ fm.
In the case of the heavy system (Pb+Pb) the picture is very different 
and collective coordinates ($\tau,\eta$) have no preference compared 
to ($t,z$). The resonances produce the long tail in the 
($\tau,\eta$)-plot which is strongest pronounced in the case of the
S+S collision. 

\begin{figure}[htp]
\centerline{\epsfysize=15cm \epsfbox{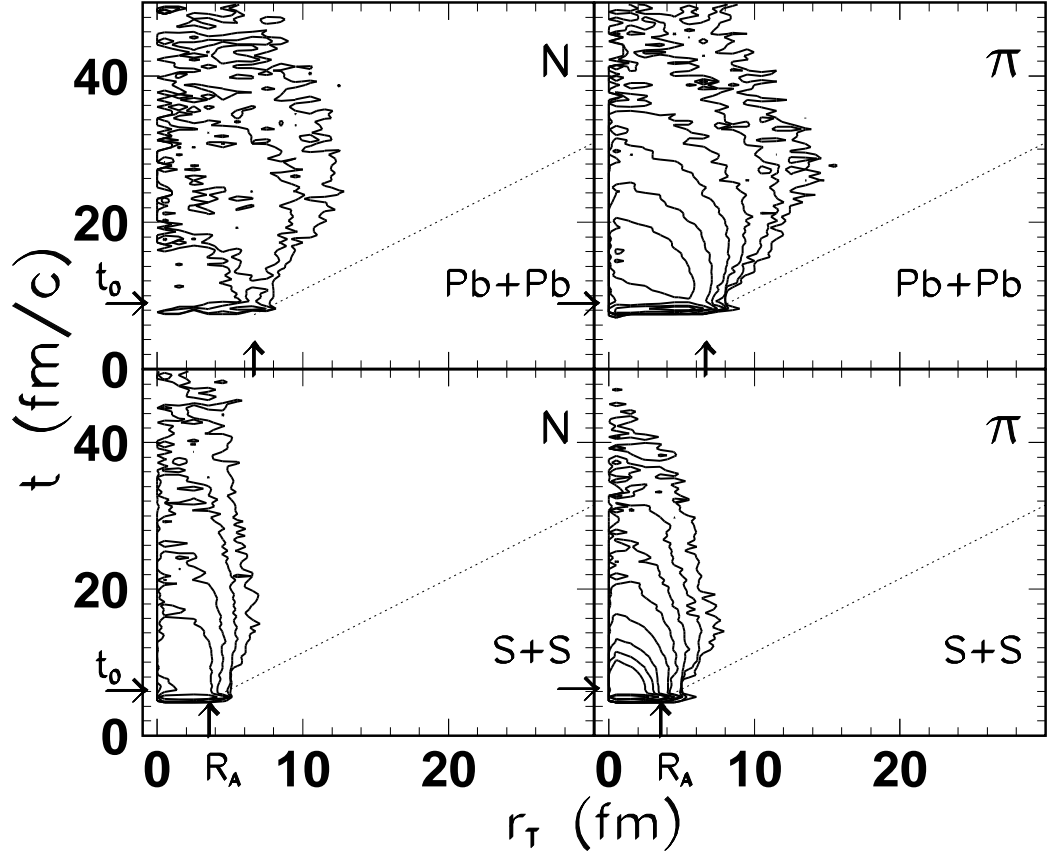}}
\caption{
$d^2N/r_Tdr_Tdt/A$ distribution of the final state hadrons over their 
last elastic and inelastic collision points in $(r_T,t)$-plane. 
Distributions are presented separately for nucleons (left panels)
and pions (right panels) produced in central S+S (lower row) and Pb+Pb 
(upper row) collisions at 160 AGeV/$c$. Contour plots correspond to
$d^2N/dtr_Tdr_T/A=$ 0.001, 0.003, 0.01, 0.03, 0.1, 0.3, 0.6, 1.0, 3.0, 
6.0 particles/(fm$^2$/$c$).
The dotted lines correspond to the line $r_T=R_A+v_{cm}(t-t_0)$.
}
\label{fig2}
\end{figure}

Figure~\ref{fig2} shows the distribution $d^2N/r_Tdr_Tdt$ of the 
emitted nucleons and pions over time, $t$, and transverse radius, 
$r_T=\sqrt{(x^2+y^2)}$. It is evidently different for nucleons and 
pions. Pions are emitted evenly from the whole volume of the reaction.
As one can see in Fig.~\ref{fig2} the emission from the inner part of 
the Pb+Pb nucleon system is strongly suppressed at times 
$t-t_0=18$ fm/$c$. This is due to the formation time effect at the 
initial stage of the reaction and longer mean free path of pions (note 
that $\sigma _{\pi N}\cong 20$mb, while $\sigma _{NN}\cong 40$mb).

Nucleons, because of the small formation time and shorter mean free 
path leave the system initially from the surface region only,
$r_T\approx R_A$. Due to the transverse flow the maximum of the 
distribution is moving to the larger $r_T$ at later times. 
For S+S this effect is less pronounced. Only in the case of nucleons
the picture is closer to the conventional picture of the freeze-out
with a relatively narrow emission region.
In comparison with Au+Au and S+S collisions at AGS energies a strong 
collective transverse expansion of hadronic matter is observed,
particularly for nucleons in Pb+Pb collisions.

In the projection one can see the cross section of the freeze-out 
hypersurface. For nucleons in Pb+Pb reaction the surface around $R_A$ 
initially is only 1 - 2 fm thin, later it becomes timelike and widens 
to $\Delta t \approx 10$ fm. If we take a narrow $z$- interval instead 
of integrating over the whole $z$-axis, the timelike thickness of the 
front would be even smaller. It means that the assumption of a
freeze-out surface {\it in this case\/} is not unreasonable.

\begin{figure}[htp]
\centerline{\epsfysize=15cm \epsfbox{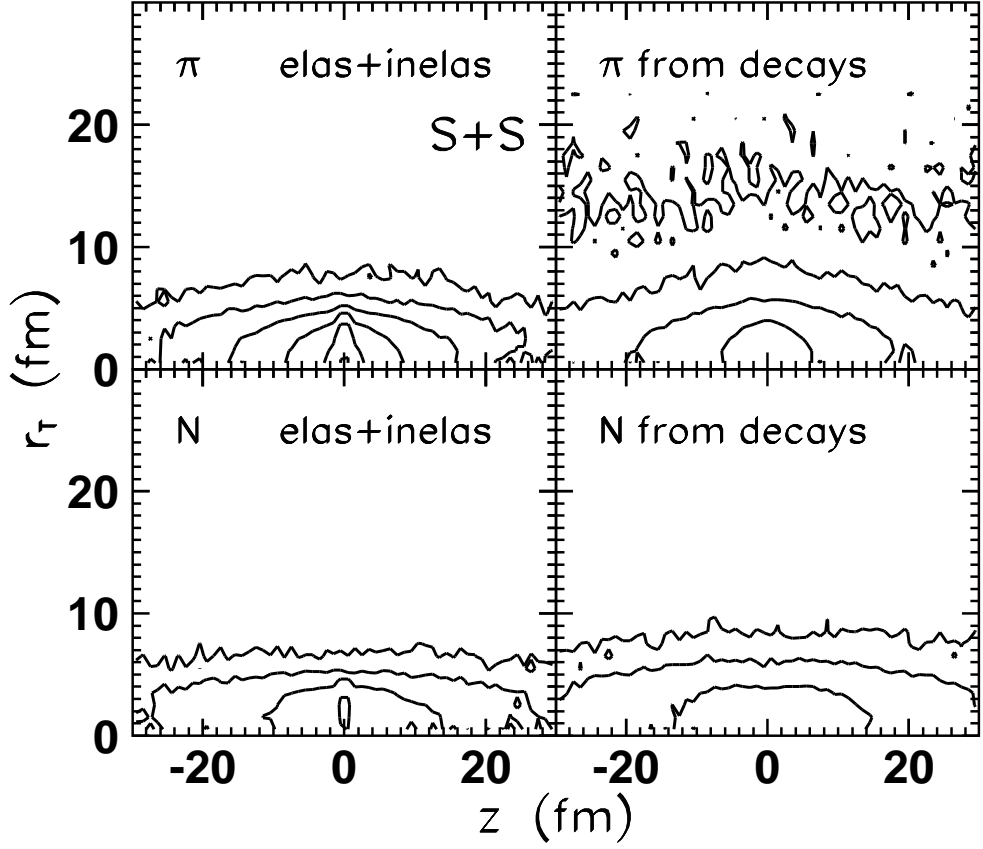}}
\caption{
$d^2N/r_Tdr_T dz$ distribution of the final state hadrons, produced 
in central S+S collisions at 160 AGeV/$c$, over their last elastic 
and inelastic collision points in $(z,r_T)$-plane. Distributions are 
presented separately for nucleons (lower panels) and pions (upper 
panels). Contour plots correspond to $d^2N/dzr_Tdr_T=$ 0.001, 0.01, 
0.1, 0.5, and 5.0 particles/fm$^3$. The maximum value of the
distribution is 0.5 fm$^{-3}$ for nucleons 
and 15.5 fm$^{-3}$ for pions.  
}
\label{fig3}
\end{figure}

Correlations between transverse and longitudinal coordinates 
$(z,r_T)$, of last collision points, $d^2N/dzr_Tdr_T$, are shown in 
Fig.~\ref{fig3} for hadrons produced separately in elastic and
inelastic interactions and in decays of resonances in S+S collisions.
It is easy to see that pions are emitted from the central zone: 
$z\approx 0, r_T\approx 0$. For nucleons the maximum of the 
$d^2N/dz r_Tdr_T$-distribution is shifted to the surface of the 
nucleus: $z=0, r_T\approx r_A$. This confirms the fact that nucleons 
at the beginning of the reaction are coming from the nuclear surface. 
Note also, that for all species longitudinal expansion is stronger 
than the transverse one, which spreads up to $r_T\approx 10$ fm.

\subsection{Distributions of the emitting sources}
\label{s3sub2}

Sometimes it is useful to consider one-dimensional distributions 
${dN}/{dz}$, ${dN}/{r_Tdr_T}$ and $dN/dt$ instead of ${d^2N}/{dzdt}$ 
and ${d^2N}/{r_Tdr_Tdt}$. 

\begin{figure}[htp]
\centerline{\epsfysize=15cm \epsfbox{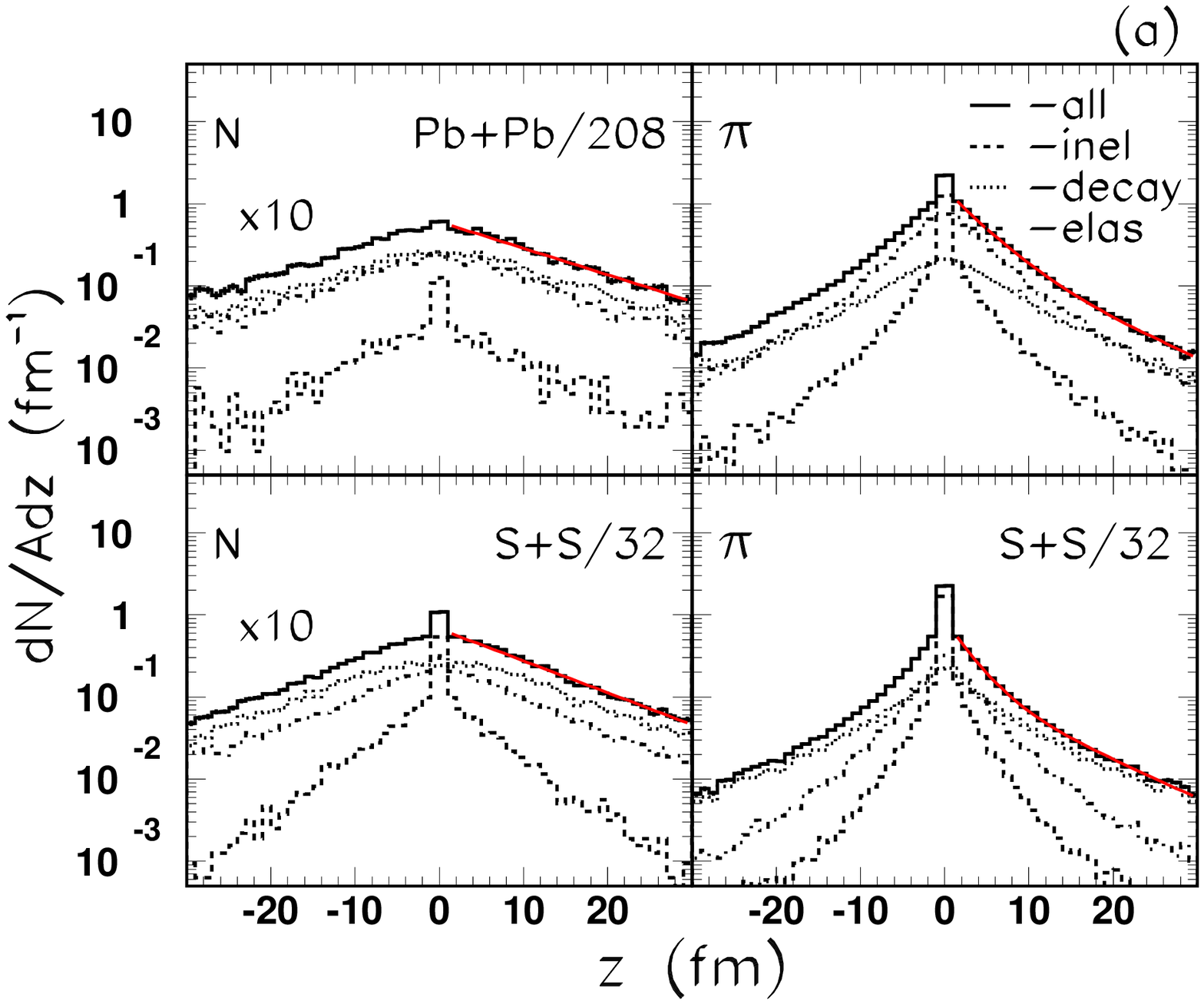}}
\caption{
$dN/A dz$ {\bf (a)} and $dN/A r_Tdr_T$ {\bf (b)} distributions
of the final state pions (right panels) and nucleons (left panels)
over their last interaction coordinates normalized to the nuclear 
mass number $A=32$ and $208$ for $S+S$ (lower row) and $Pb+Pb$ (upper 
row) collisions, correspondingly. Distributions for nucleons are 
scaled up by an order of magnitude. Dashed, dash-dotted, dotted lines 
and solid histograms correspond to the inelastic and elastic 
collisions, resonance decays and the overall sum, respectively. The 
solid lines correspond to the fit of the total distributions to the 
sum of two exponentials $C_1*\exp{(-z/R_{L1})}+C_2*\exp{(-z/R_{L2})}$ 
for pions and to one exponential, $C*\exp{(-z/R_L)}$, for nucleons.
The values of parameters are listed in Table \protect \ref{tab2}.
}
\centerline{\epsfysize=15cm \epsfbox{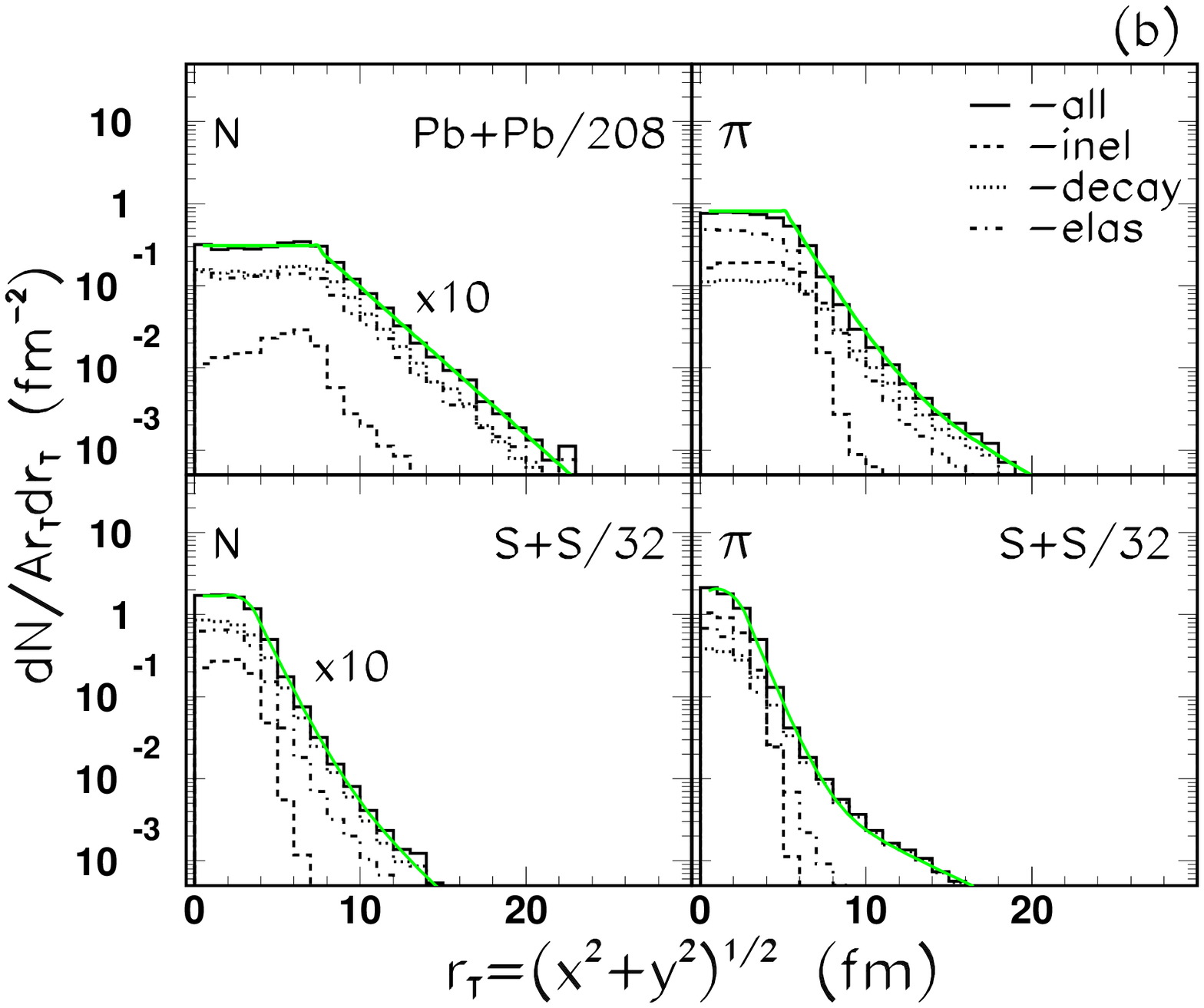}}
\label{fig4}
\end{figure}

Figure~\ref{fig4}(a)-(b) shows the time integrated distributions of 
emitted nucleons and pions over their longitudinal, $z$, and 
transverse, $r_T$, coordinates. The striking feature of these 
distributions is their nontrivial shape, which is neither Gaussian nor 
exponential.

This is particularly clear for pions which show a sharp peak in 
$dN/dz$ distributions at $z=0$ in both systems. The pion distributions 
in Pb+Pb (S+S) reactions can be fitted to the sum of two exponentials,
$C_1*\exp{(-z/R_{L1})}+C_2*\exp{(-z/R_{L2})}$. 
The $z$-distribution of nucleons in both reactions is close to the 
exponential. The parameters of the fits are listed in 
Table~\ref{tab2}.

For all reactions the $dN/A r_Tdr_T$ distributions are flat within 
the radius of the colliding nuclei. In Pb+Pb the nucleon emission is 
even slightly peaked at $r_T=R_{Pb}$, as seen in Fig.~\ref{fig4}(b).
At larger transverse radii distributions for pions in Pb+Pb (S+S) 
collisions and for nucleons in S+S collisions are nicely fitted
by the sum of two exponentials 
$C_1*\exp{(-r_T/R_{C})}+C_2*\exp{(-r_T/R_{H})}$.
The long tail in the distributions is due to the decays of long-lived 
resonances, which become significant at $r_T=15$ (11) fm for Pb+Pb 
(S+S) collisions. For nucleons in Pb+Pb the distribution is very close 
to the single exponential, $C*\exp{(-z/R_C)}$ (see Table~\ref{tab2}).
This is because of narrower emission region for nucleons in Pb+Pb 
collisions (see Fig.~\ref{fig2}).

Even for small $\Delta t$ intervals the $dN/r_Tdr_T$-distributions 
over $r_T$ are not similar to Gaussians. They become broader with 
growing time due to the transverse expansion. At later times their 
shape changes drastically because of the dominant contribution of the 
resonance decays.

\begin{figure}[htp]
\centerline{\epsfysize=15cm \epsfbox{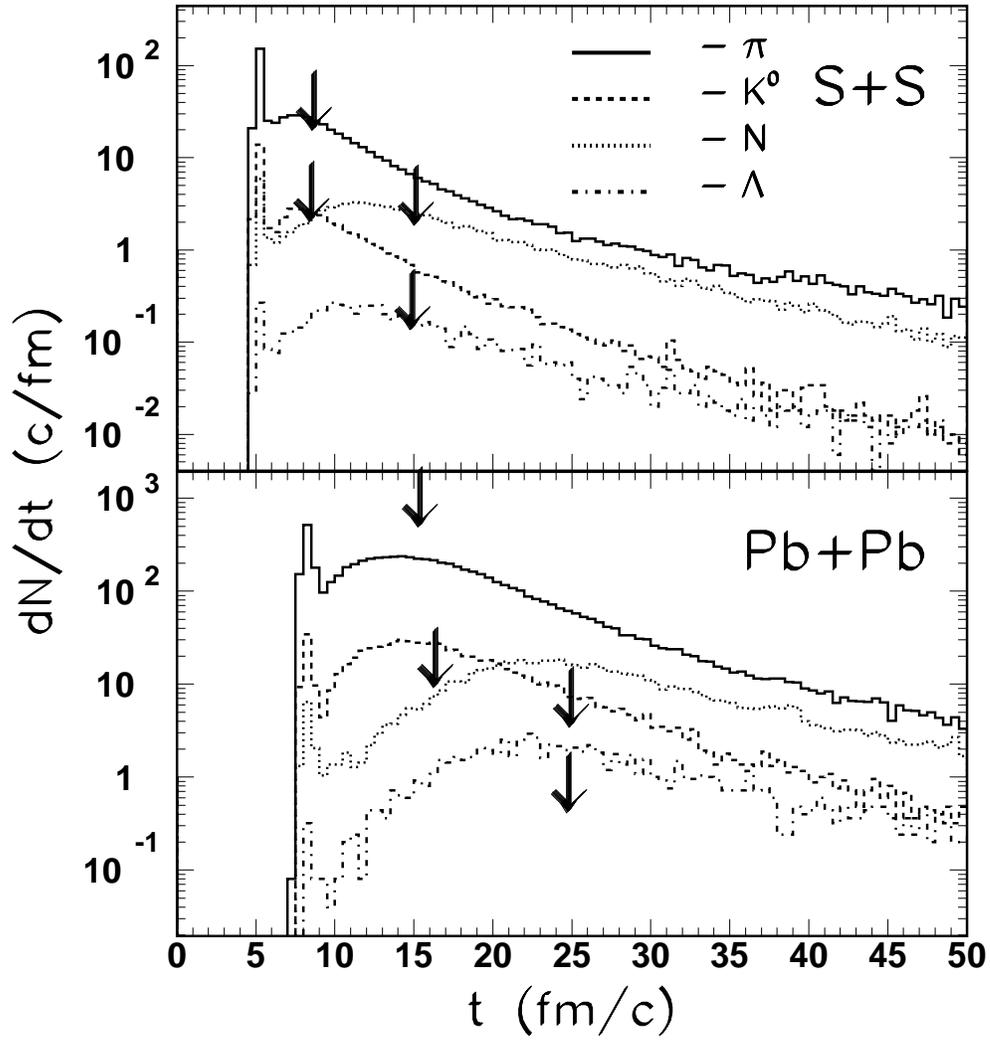}}
\caption{
$dN/dt$-distribution of the particles over their last collision time, 
$t$, for kaons (dashed), pions (solid), nucleons (dotted) and lambdas 
(dash-dotted histograms) for S+S (upper row) and Pb+Pb (lower row) 
collisions. The vertical arrows correspond to the average emission 
times of the species.
}
\label{fig5}
\end{figure}

Figures~\ref{fig4} and \ref{fig5} show separately the contributions of 
the hadrons which go to the detector after their last inelastic and  
elastic collision or a resonance decay. The inelastic collision points 
are located in a narrow space-time region, close to the beginning of 
the reaction. These particles leave the contact zone before the nuclei 
overlap strongly. The last elastic collisions are spread in much wider 
space-time domain, particularly, in Pb+Pb collisions. These particles 
reflect the expansion of the hadronic matter produced in the reaction.
Most of the inelastically produced particles do not leave the system 
but suffer subsequent elastic collisions. Elastic and inelastic 
freeze-outs are well separated. The time interval where inelastic 
collisions are frequent enough is probably too short to speak about 
chemical equilibrium. 

A large group of particles reaches detectors  after the resonance 
decays, e.g., $\Delta\rightarrow \pi+N$, $\rho\rightarrow 2\pi$, 
$\omega\rightarrow3\pi$. These particles become dominant at $t=25$ 
(7.5) fm/$c$ after the first collision in the Pb+Pb (S+S) reactions. 
The maximum of the distribution for resonance decays is shifted to the 
later times compared to the distribution of the particles decoupled 
after the last elastic or inelastic collisions. 
The distribution for resonance decays is much broader than that for 
inelastic collisions, but much narrower than the one for elastic 
collisions both in Pb+Pb and in S+S reactions.  

In S+S and Pb+Pb collisions the $dN/dt$ distributions presented in 
Fig.~\ref{fig5} have a sharp peak corresponding to the first (which is 
also the last) inelastic collision. Subsequent elastic rescatterings 
or decays spread out the distribution to the later times. In Pb+Pb 
collisions the contribution of initial interactions is almost entirely 
washed out for nucleons,  because of the large number of 
rescatterings. But even in this case the distributions are wide with 
maxima shifted to later times.

\subsection{Sequential freeze-out}
\label{s3sub3}

Let us now consider in more details the time distributions for the 
different hadron species shown in Fig.~\ref{fig5}. In both systems 
there is a noticeable difference between the meson and baryon groups 
of particles. The QGSM predicts that kaons and pions decouple earlier 
than nucleons and lambdas and approximately at the same times 
$\langle t^{mes} \rangle \approx 17.5$ (10) fm/$c$ and $\langle 
t^{bar} \rangle \approx 26.5$ (17.5) fm/$c$ for the Pb+Pb (S+S) 
reaction. The width of $dN/dt$- distributions for mesons are narrower 
than that for baryons: $\Delta t^{mes}\approx 7.5$ (6.5) fm/$c$ and
$\Delta t^{bar}\approx 8.5$ (9.0) fm/$c$  in Pb+Pb (S+S) case.
For $K$'s and $\Lambda$'s the width is slightly smaller than the 
width for pions and nucleons, respectively.  
At the last stages of the reaction the $dN/dt-$ distributions  
for nucleons and pions are determined mainly by the resonance decays
$\Delta \rightarrow \pi + N$, while the width of the distributions  
of kaons and lambdas is determined by the elastic collisions. 
At this stage pions and nucleons (as well as kaons and  lambdas) have 
the same decoupling times and the slopes of $dN/dt$-distributions.          

Therefore, our microscopic model clearly shows that there is no 
unique freeze-out time for different hadrons at SPS energies. In fact,
the particles are emitted continuously. The mesons are emitted by 
about 10(6) fm/$c$ earlier than baryons in Pb+Pb (S+S) collisions at 
160 AGeV/c. This conclusion is valid also for the particles emitted in 
a certain rapidity interval, particularly at central rapidities,
$|y| \leq 1$.

\section{Phase-space correlations} 
\label{sec4}

In contrast to experiment, which is dealing only with the momenta of 
the final state particles, the microscopic model provides the full 
information about the produced particles, such as coordinates, times 
and momenta. Below we study the global correlations between the 
momentum distributions of hadrons and the space and time 
characteristics of the emission source, such as the rapidity-time and 
transverse momentum-time distributions. This information is especially 
important for the theoretical interpretation of interferometric 
measurements.

\subsection{Transverse mass and time}
\label{s4sub1}

\begin{figure}[htp]
\centerline{\epsfysize=15cm \epsfbox{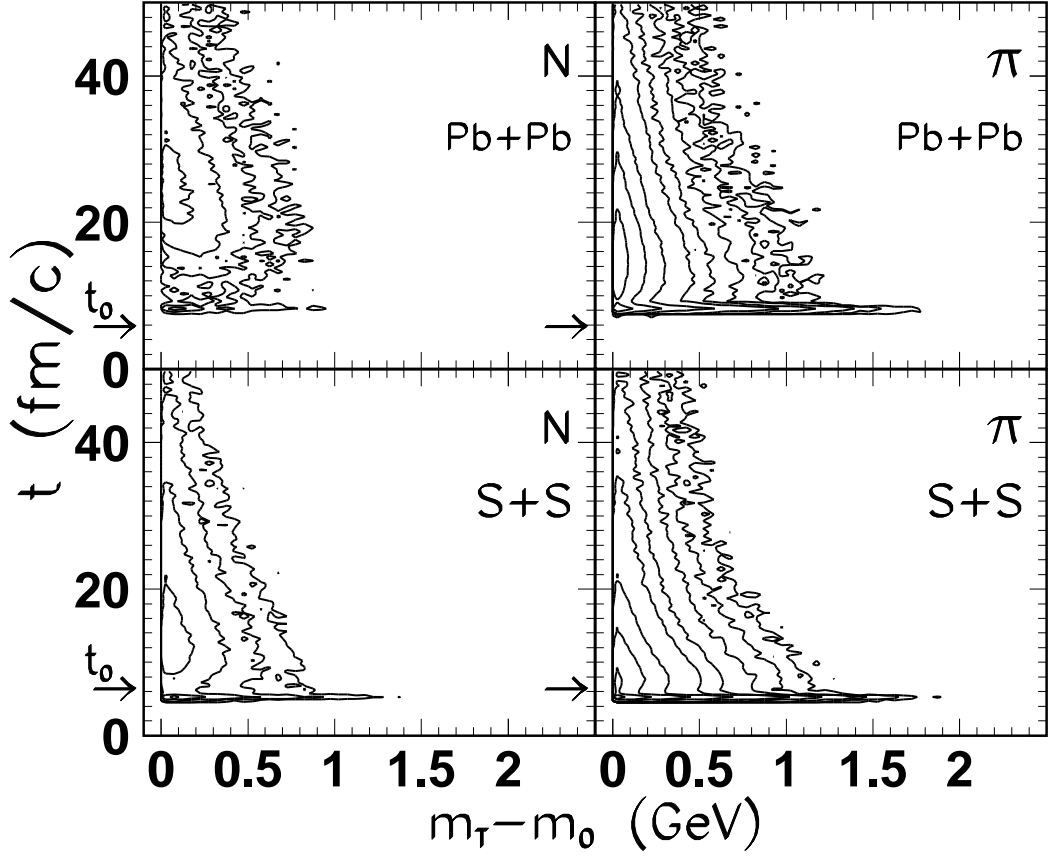}}
\caption{
$d^2N/m_Tdm_Tdt/A$-distribution of the final state hadrons over 
transverse mass, $m_T$, and the emission time, $t$, for nucleons
(left panels) and pions (right panels). The yields are divided by 
the mass number of the colliding nuclei, $A=32$ for S+S (lower row) 
and $A=208$ for Pb+Pb (upper row). Contour plots correspond to
$d^2N/m_Tdm_Tdt/A=$0.003, 0.01, 0.03, 0.1, 0.25, 1.0, 3.0, 10.0
particles/(fm$^2$/$c$). 
}
\label{fig6}
\end{figure}

Figure~\ref{fig6} shows the contours of the $d^2N/m_Tdm_Tdt$ 
distribution of the final state hadrons in the $(m_T,t)$ plane. Here 
$m_T=\sqrt{p_T^2 + m_0^2}$ is the transverse mass of a particle with 
rest mass, $m_0$. One can see the difference between pion and nucleon 
emission in the case  of heavy nuclei. The pions with large transverse 
momenta are emitted only at the initial stages of the S+S and Pb+Pb 
reactions. They are produced in inelastic primary $NN$ collisions. In 
contrast to pions the nucleons with maximal transverse momenta in 
Pb+Pb collisions are coming from the intermediate times 
$(t-t_0)=12-14$ fm/$c$. Soft hadrons are emitted during the whole 
evolution time. The maximum of the emission rate in S+S corresponds to 
the initial time of the reaction, while in Pb+Pb it is shifted to 
about $t-t_0=14-18$ fm/$c$ because of many rescatterings. With growing 
time the $m_T$-spectra become gradually softer, that can be 
interpreted as the cooling of the expanding hadronic matter.

We see that in the case of nucleons in Pb+Pb collisions the 
contribution of the particles emitted after the first interaction is 
completely washed out. The transverse momenta are generated to large 
extend by multiple rescatterings. It is very likely, therefore, that 
in Pb+Pb collisions we indeed are dealing with more or less 
thermalized source.

\begin{figure}[htp]
\centerline{\epsfysize=15cm \epsfbox{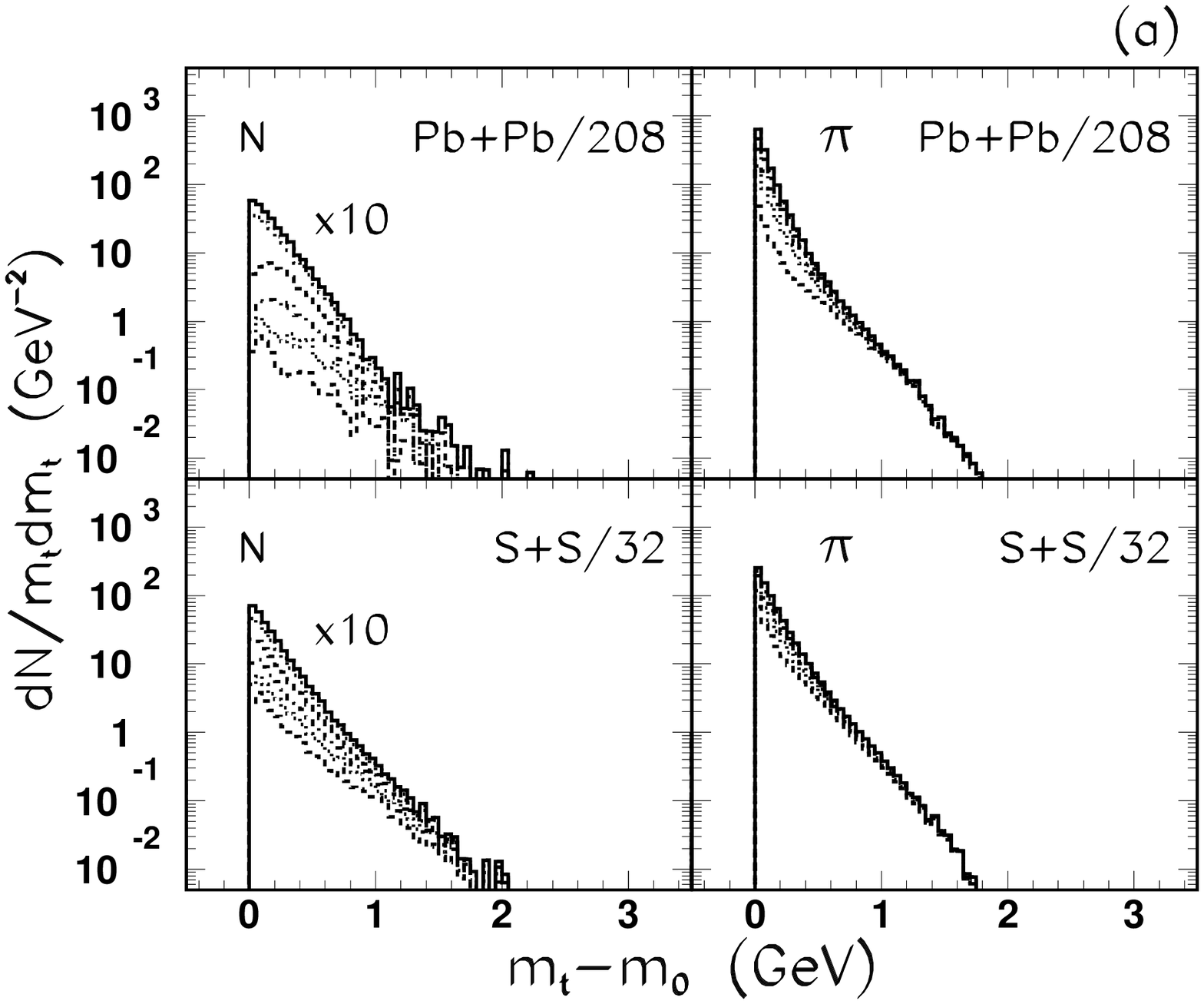}}
\caption{
{\bf (a)}: 
$dN/A m_Tdm_T$- distributions of the final state pions (right panels) 
and nucleons (left panels) integrated over emission times from $t=0$ 
to $t=6-8-10-13-20-50$ fm/$c$ for S+S (lower row) and to 
$t=9-13-16-20-30-50$ fm/$c$ for Pb+Pb (upper row) collisions.\\
{\bf (b)}: 
$dN/A m_Tdm_T$  distributions of the final state pions (right panels) 
and nucleons (left panels) over their transverse mass, $m_T$, produced 
in inelastic collisions (dashed curves), in resonance decays (dotted
curves) or elastic collisions (dash-dotted curves), and for sum of 
them (solid curve).
}
\centerline{\epsfysize=15cm \epsfbox{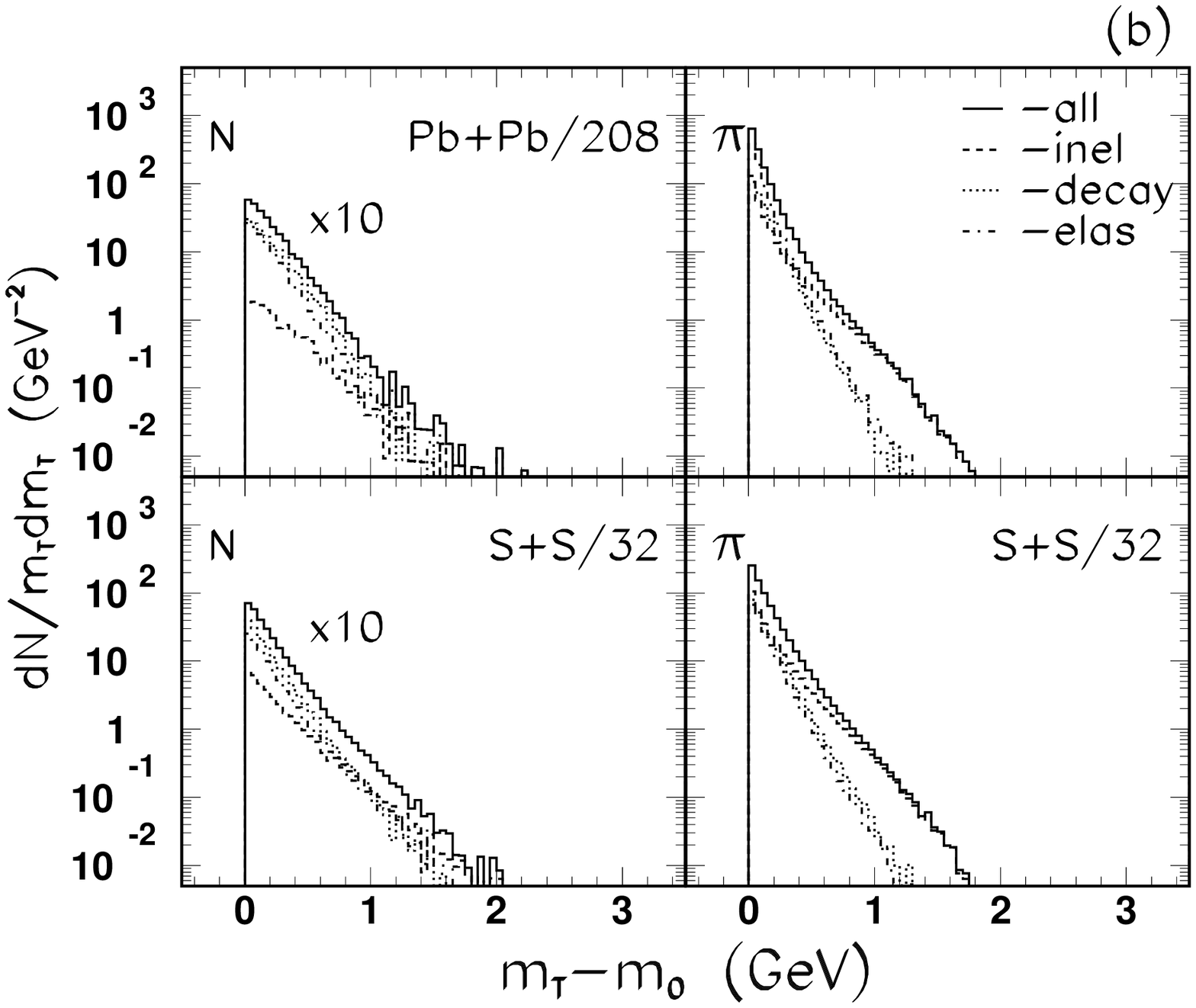}}
\label{fig7}
\end{figure}

The time evolution of the $m_T$ spectra is presented in 
Fig.~\ref{fig7}(a). The spectra of pions are the widest at the 
beginning of the reaction (small $t$) and become steeper at large 
$t$. The particles with different transverse momenta are emitted 
from regions with different characteristic size. Since the system is 
expanding and, therefore, it is smaller at early times, it is
tempting to conjecture that particles with large $p_t$ come from a 
region with relatively small longitudinal and transverse size.

The final $m_T$-spectra, shown in Fig.~\ref{fig7}(b), are composed of 
hadrons, produced in inelastic and elastic collisions, and resonance 
decays. Elastic rescatterings make the particle spectra softer. The 
products of resonance decays populate the soft parts of the spectra 
while the collective flow, caused mostly by elastic collisions, leads
to the broadening of the spectra.
The shoulder-like structure appears in the spectra at large transverse 
momenta. The final transverse momentum distributions are rather 
complex and cannot be reproduced by a single thermal source model.

\subsection{Rapidity and time}
\label{s4sub2}

\begin{figure}[htp]
\centerline{\epsfysize=15cm \epsfbox{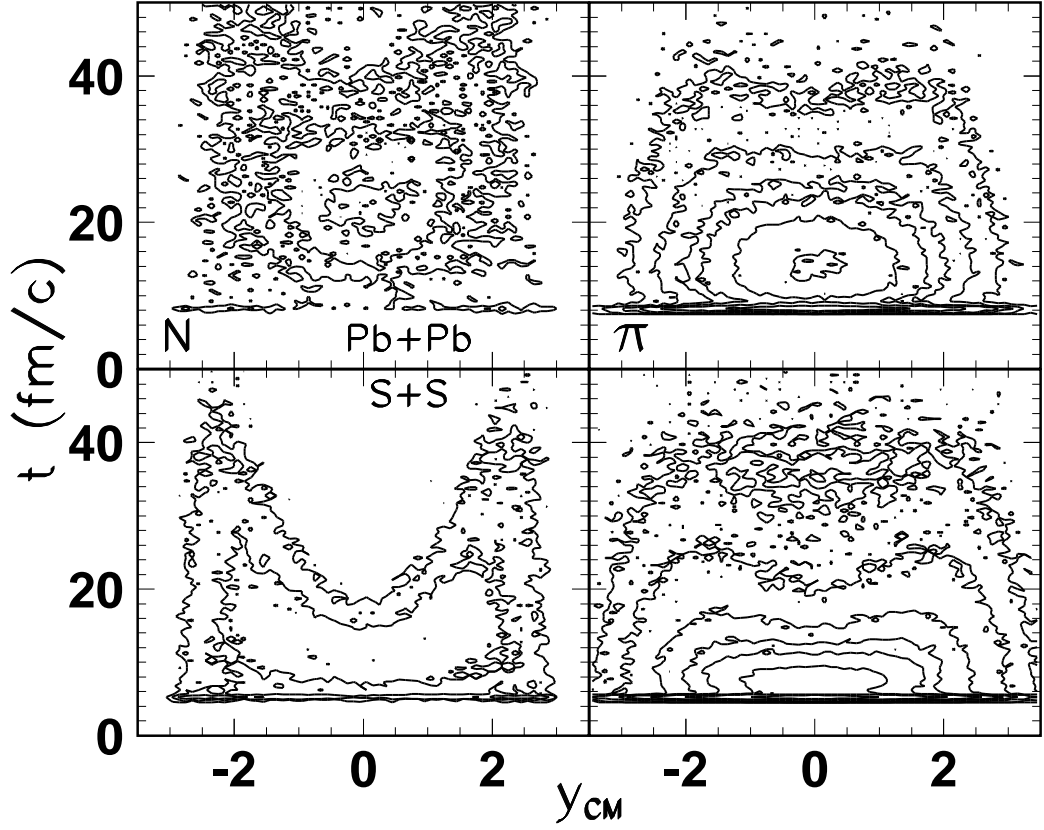}}
\caption{
$d^2N/dy_{cm}dt/A$ distribution of the final state hadrons over 
rapidity, $y_{cm}$, and emission time, $t$. Notations are the same as 
in Fig. \protect \ref{fig2}. Contour plots correspond to
$d^2N/y_{cm}dt/A=$0.003, 0.01, 0.033, 0.07, 0.11, 0.2, 0.4, 1.0, 3.0, 
10.0 particles/(fm/$c$).
}
\label{fig8}
\end{figure}

Figure~\ref{fig8} shows the contours of the $d^2N/dydt$ distribution 
in the $(y,t)$ plane. The maximum of the distribution in Pb+Pb 
collisions is shifted to the same times $t\approx 12-14$ fm/$c$ after 
the beginning of the reaction as in the $(m_T,t)$-distribution. In S+S 
the nucleons with large rapidities are produced at initial times, as 
well as particles with large $p_T$. This effect is less pronounced for 
Pb+Pb. At the latest times one can see a significant difference 
between the two reactions considered. In the Pb+Pb collision the 
rapidity spectra are more equilibrated as in the S+S system. At the 
initial stage of both reactions the nucleon rapidity spectra have a 
characteristic two-hump structure like in $NN$ interactions. 
The initial Fermi motion makes the rapidity spectra of the nucleons 
wider. At intermediate times the central part of the distribution
is growing gradually with time and reaches the maximum at 12(6) 
fm/$c$ after the beginning of the Pb+Pb (S+S) reaction. 
As one can see from Fig.~\ref{fig8} most nucleons are emitted in the 
central rapidity window $|y| \leq 1$. In Pb+Pb collisions this part 
becomes dominant at times of about 14 fm/$c$ after the beginning of 
the reaction, while in S+S its contribution remains small. The 
evolution of the spectra reflects clearly the different degree of 
stopping reached in the two reactions. The rapidity spectra in Pb+Pb 
collisions look like emission from a thermal source for 
$t-t_0>16$ fm/$c$. At SPS energies in both reactions one can see 
higher transparency of the nuclear matter as compared with AGS 
energies \cite{Freeze,Freeze2}.

\begin{figure}[htp]
\centerline{\epsfysize=15cm \epsfbox{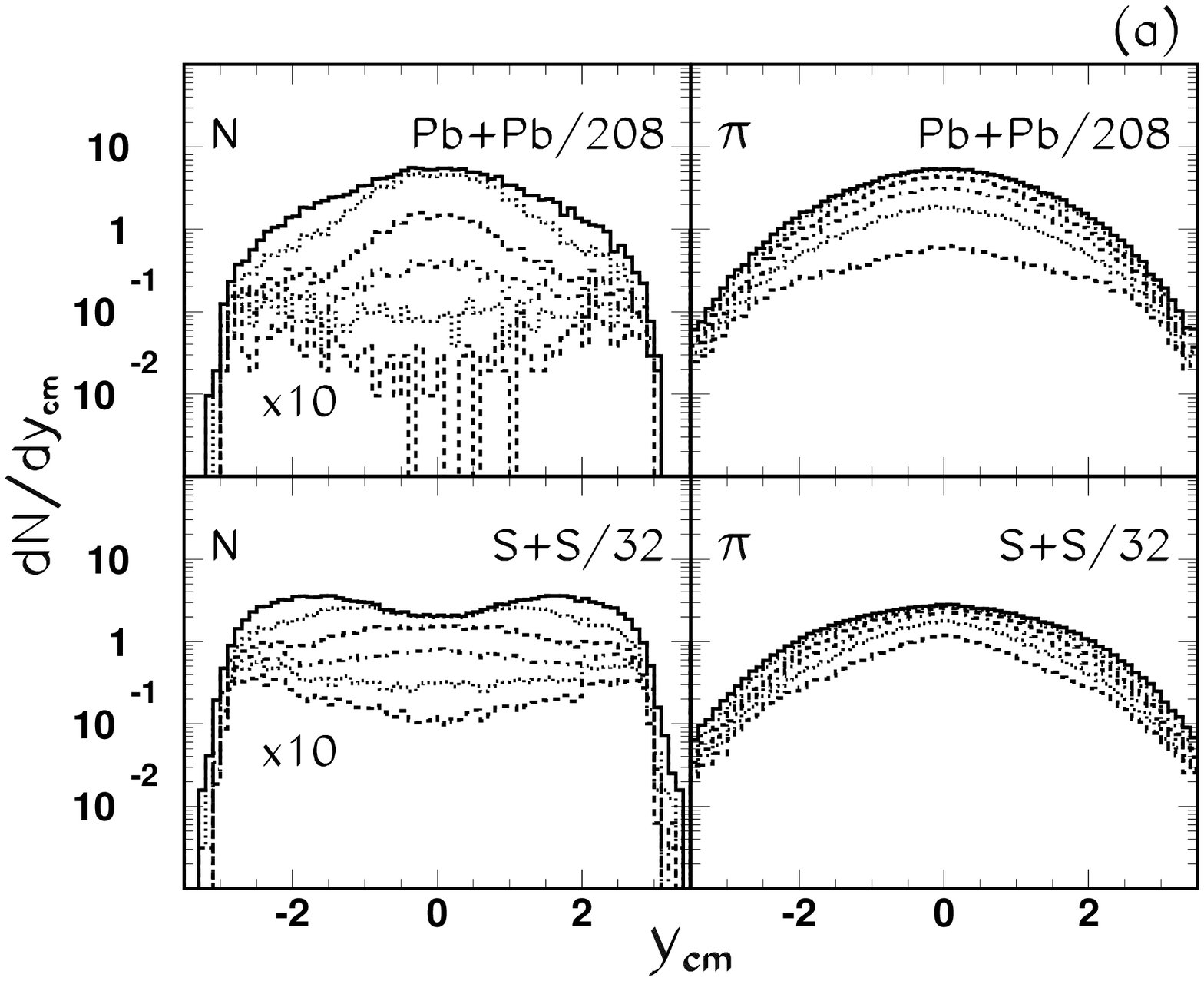}}
\caption{
{\bf (a)}:
The same as Fig. \protect \ref{fig7}(a) but for
$dN/A dy_{cm}$  distributions of the final state pions (right panels) 
and nucleons (left panels) over their rapidities, $y_{cm}$. \\
{\bf (b)}:
The same as Fig. \protect \ref{fig7}(b) but for
$dN/A dy_{cm}$  distributions of the final state pions (right panels) 
and nucleons (left panels) over their rapidities, $y_{cm}$.  
}
\centerline{\epsfysize=15cm \epsfbox{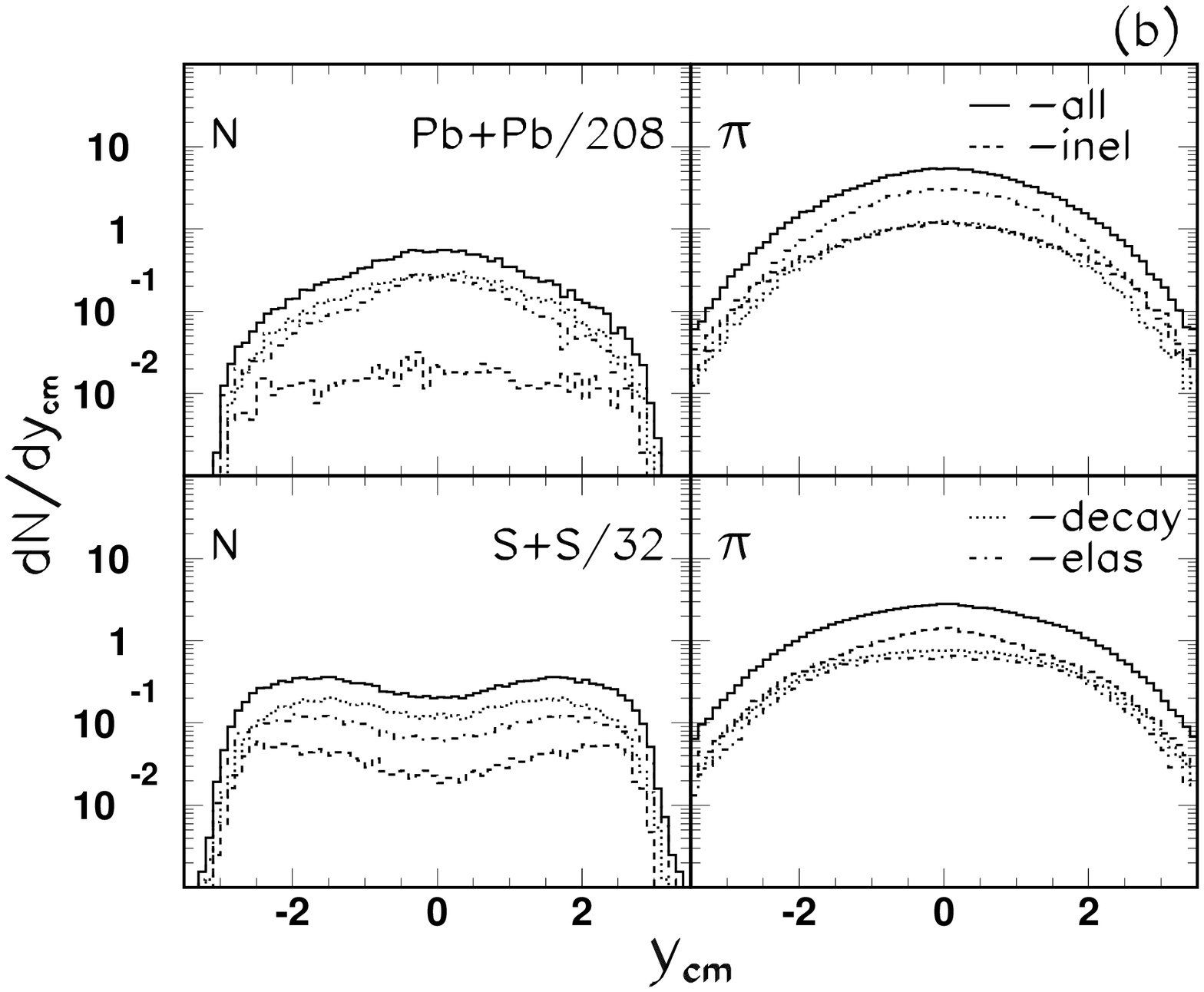}}
\label{fig9}
\end{figure}

The decreasing width of the rapidity spectra with growing time
is in agreement with our previous conclusion on the cooling of the 
system. In both reactions the two-hump structure of the spectra 
survives even at latest times reflecting the residual longitudinal 
motion of the baryons. These trends are clearly seen also in 
Fig.~\ref{fig9}(a), where the rapidity spectra integrated over $t$
are shown for different times. In contrast to AGS energies the 
significant changes in nucleon spectra are seen even at the latest 
stages of the reaction, when the longitudinal expansion makes it 
broader in the Pb+Pb case and creates a dip at midrapidity in S+S 
collisions.

The final rapidity spectrum of nucleons has a bell-like shape in 
Pb+Pb collisions and two-hump shape in S+S collisions. As expected, 
there is a large difference in the behavior of the rapidity spectra 
of pions and nucleons. For pions the variation with time of the 
spectra is less pronounced because even in $NN$ collisions they have 
the bell-like shape, similar to thermal spectra. The final spectra for 
pions and nucleons are close to each other.

The contributions of different reaction channels (elastic, inelastic, 
decays) are shown in Fig.~\ref{fig9}(b). As was mentioned before, in 
both reactions the decays give the most significant contribution at 
the latest times. In both reactions half of the nucleons are coming 
from the resonance decays, 40\% from the elastic collisions and about 
10\% from inelastic collisions. Third of the pions are coming from 
the resonance decays, 23\% (40\%) from the inelastic collisions and 
52\% (23\%) from elastic collisions in Pb+Pb (S+S) reactions at SPS.

The observed picture has an important implication for particle 
interferometry \cite{KP74,MS88,KG86,CSH95,AS96,CL96}. According to 
QGSM, nucleons with large $|y|$ show small longitudinal size of the 
emitting source while particles from the midrapidity region indicate 
a large longitudinal size and life time of the source.

\section{Conclusions}
\label{sec5}

From the analysis of various space-time and phase space particle
distributions, obtained within the QGSM model for light (S+S) and 
heavy (Pb+Pb) systems of colliding nuclei at SPS energies, the 
following conclusions may be drawn.
The system of final particles in heavy ion collision can be 
represented as a core and a halo. The core contains the particles 
which are still in evolution through the inelastic and elastic 
collisions. The halo is represented by particles which are already 
decoupled from the system and move to the detectors.

Microscopic models like QGSM or UrQMD cannot in principle give a 
sharp freeze-out due to the lack of attractive forces which keep the 
particles together. To get a sharp freeze-out it is necessary to 
have some glue mechanism, like attractive mean fields, enhanced cross 
sections or rapidly hadronizing Quark Gluon Plasma.

The shapes of the emitting sources are far from Gaussians. In addition,
the $\tau$- scaling, which is often used in the parameterizations, is 
not confirmed by the model calculations for heavy systems like Pb+Pb. 
More realistic source shapes and freeze-out criteria should be used in 
the analysis of HBT interferometric data. 

These results are supported by the results of recent paper 
\cite{MaSu99}, in which the importance of freeze-out models for the
HBT analysis has been investigated. In this article the HBT
correlators of both identical and non-identical pions are shown to
depend strongly on the pion production scenario. Also, a more 
realistic model of freeze-out, in which the dynamics of the system
is driven by binary collisions, has been elaborated. Note, that the
case of heavy ion collisions at relativistic energies simulated by 
QGSM, with the pions (as well as other particles) coming both from
elastic and inelastic collisions and from decays of resonances, is
much more complex compared to the elastic freeze-out scenario and
requires further theoretical analysis.

Our main conclusion is that the Quark Gluon String Model predicts a 
continuous emission of particles, starting almost from the beginning 
of the reaction. One can consider this picture as a limiting case of 
the dynamics of relativistic heavy ion collisions. The hydrodynamical 
sharp freeze-out presents another idealized limit (for recent review 
see \cite{HJ99,BG99,Cs99} and references therein).
We believe that the truth is somewhere between. 
Certainly, the rich experience of microscopic calculations should be
employed for constructing  more sophisticated macroscopic models.

\section{Acknowledgments}

We are grateful to H. Heiselberg and H. St{\"o}cker  
for fruitful discussions. This work is supported in part by Danish 
Natural Science Research Council, A.v.Humboldt Stiftung, the 
EU-INTAS-grant 94-3405, and the Bundesministerium f\"ur Bildung and
Forschung (BMBF).
L.B. thanks the Niels Bohr Institute and NORDITA for kind hospitality 
and financial support. 
I.N.M. thanks the Carlsberg Foundation for the financial support.

\newpage

\newpage
\mediumtext

\begin{table}
\caption{
The total number of the collisions, $N_{coll} = N_{inel} + N_{el}$,
resonance decays, $N_{dec}$, baryons, $B$, and mesons, $M$, produced 
in central Pb+Pb and S+S collisions at 160 AGeV/$c$, and in Au+Au and 
S+S collisions at 11.6 AGeV/$c$ at $t=50$ fm/$c$. 
}

\vspace*{1.0cm}

\begin{tabular}{ccccc}
 & \multicolumn{1}{c}{Pb+Pb} & \multicolumn{1}{c}{Au+Au } 
 & \multicolumn{1}{c}{S+S  } & \multicolumn{1}{c}{S+S }  \\ 
 & \multicolumn{1}{c}{160 AGeV/$c$} & \multicolumn{1}{c}{11.6 
    AGeV/$c$} 
 & \multicolumn{1}{c}{160 AGeV/$c$} & \multicolumn{1}{c}{11.6 
    AGeV/$c$}\\ 
\tableline
 $N_{coll}$ & 18595 & 8972    &  410   &  242     \\
 $N_{inel}$ & 6784  & 4096    &  206   &  129     \\
 $N_{el}  $ & 11811 & 4876    &  204   &  113     \\
 $N_{dec} $ & 1960  & 1207    &  116   &   58     \\ 
 $B$        & 417   & 384     &   58   &   55     \\ 
 $M$        & 3830  & 973     &  342   &   97     \\ 
\end{tabular}
\label{tab1}
\end{table}

\vspace*{1.0cm}

\begin{table}
\caption{
The parameters of the longitudinal $dN/dz$  
and transverse $dN/r_Tdr_T$ distributions of the emitting sources
obtained in the QGSM for pions and nucleons in central S+S and Pb+Pb
collisions at 160 AGeV fitted to the single exponential function,
$C*\exp {(-z/R)}$, and to the sum of two exponential functions,
$C_1*\exp {(-z/R_{1})} + C_2*\exp {(-z/R_{2})}$.   
}

\vspace*{1.0cm}

\begin{tabular}{ccccc|cccc}
&\multicolumn{4}{c|}{$dN/dz$}&\multicolumn{4}{c}{$dN/r_Tdr_T$}\\ 
\tableline
&\multicolumn{1}{c}{$C_1$}&\multicolumn{1}{c}{$R_{L1}$}&
\multicolumn{1}{c}{$C_2$}&\multicolumn{1}{c|}{$R_{L2}$}
&\multicolumn{1}{c}{$C_1$}&\multicolumn{1}{c}{$R_C$}
&\multicolumn{1}{c}{$C_2$}&\multicolumn{1}{c}{$R_H$}\\ 
\tableline
$\pi $ (S+S)   & 0.722 & 2.65    &  0.14    & 9.35  &
                 18.89 & 0.91    &  0.0183  & 4.56   \\ 
$\pi $ (Pb+Pb) & 1.25  & 3.81    &  3.74    & 3.81  &
                 33.62 & 1.33    &  0.083   & 3.86   \\ 
$ N $  (S+S)   & 0.067 & 11.1    & $ - $    & $ - $ &
                  3.37 & 1.03    &  0.020   & 2.40   \\ 
$ N $ (Pb+Pb)  & 0.060 & 13.5    & $ - $    & $ - $ & 
                  0.77 & 2.30    & $ - $    & $ - $    \\  
\end{tabular}
\label{tab2}
\end{table}

\end{document}